\DocumentMetadata{}
\documentclass[manuscript,screen,acmsmall]{acmart}
\AtBeginDocument{%
  \providecommand\BibTeX{{%
    Bib\TeX}}}

\setcopyright{acmcopyright}
\copyrightyear{2024}
\acmYear{2024}
\acmDOI{XXXXXXX.XXXXXXX}

\acmJournal{TOCHI}
\acmVolume{XX}
\acmNumber{XX}
\acmArticle{XX}
\acmMonth{XX}
\acmConference[CSCW'25]{Make sure to enter the correct
  conference title from your rights confirmation emai}{November 
  2025}{Woodstock, NY}
\acmISBN{978-1-4503-XXXX-X/18/06}

\begin{document}

\title{Kintsugi-Inspired Design: \\ Communicatively Reconstructing Identities Online After Trauma}

\author{Casey Randazzo}
\affiliation{%
  \institution{Rutgers University}
  \city{New Brunswick,NJ}
  \country{US}}
\email{cer124@rutgers.edu}

\author{Tawfiq Ammari}
\affiliation{%
  \institution{Rutgers University}
  \city{New Brunswick,NJ}
  \country{US}}
\email{tawfiq.ammari@rutgers.edu}



\begin{abstract}
Trauma can disrupt one's sense of self and mental well-being, leading survivors to reconstruct their identities in online communities. Drawing from 30 in-depth interviews, we present a sociotechnical process model that illustrates the mechanisms of online identity reconstruction and the pathways to integration. We introduce the concept of fractured identities, reflecting the enduring impact of trauma on one's self-concept. We also conceptualize grief bubbles—algorithmic recommendations that concentrate on the traumatic aspects of users' identities, hindering identity integration. Our findings challenge the assumptions of Goffman's dramaturgy framework and extend Hecht's communication theory of identity, demonstrating how personal-enacted identity gaps manifest within online spaces. We also discuss how trauma-informed online communities can reflect the principles of Kintsugi, the Japanese art of repairing broken pottery with gold, symbolizing a renewed sense of self, before concluding with design recommendations. *Trigger warnings: Intimate partner violence, addiction, mental illness, sexual abuse, psychological maltreatment, violence, and incarceration.\end{abstract}





\received{2 July 2024}
\maketitle

\begin{flushright}
``\textit{The world breaks everyone and afterwards many are strong at the broken places.}''

—Ernest Hemingway
\end{flushright}

\section{INTRODUCTION}

In the Japanese art of Kintsugi, broken pottery is repaired using gold lacquer, highlighting their 
cracks rather than hiding them. This technique transforms fractures into lines of beauty, 
symbolizing resilience through imperfection \cite{koren1994wabi}. Similarly, trauma survivors 
often mend their fractured identities \cite{janoff-bulman_shattered_2002} by weaving traumatic experiences into a renewed sense of self \cite{herman1992}. Given the prevalence of trauma (i.e., an emotional response to distressing events \cite{kessler2013posttraumatic, norris1992epidemiology}), affecting 70\% of adults worldwide \cite{benjet_epidemiology_2016}, Chen et al. \cite{chen_trauma-informed_2022} estimated that the majority of technology users are trauma survivors. Thus, this study examines how identity reconstruction processes unfold in online communities, where platform design can widen identity fractures or, like gold lacquer in Kintsugi, help survivors redefine their sense of self.  

Researchers have demonstrated that online spaces can facilitate 
aspects of identity reconstruction during life transitions \cite{Haimsonoli2018, Andalibigarcia2021, ammari_crafting_2017, andalibi_announcing_2018, haimson_trans_2020, Semaanetal2016}.\footnote{Examples of life transitions can include becoming parents or coming out as LGBTQIA+ \cite{ammari_crafting_2017, haimson_trans_2020}.} Such studies suggest that online communities are more than passive spaces, and instead can be tools for strengthening identities during pivotal life changes \cite{Semaanetal2017, craig2014you}. However, transitional experiences are not inherently traumatic and often present a gradual path toward change \cite{ammari_accessing_2014, ammari_thanks_2016, ammari_networked_2015}. By contrast, trauma is disruptive, often abrupt, and entails deeper emotional and 
psychological consequence \cite{herman1992, janoff-bulman_shattered_2002}. Accordingly, trauma-informed design principles \cite{chen_trauma-informed_2022}, adapted from the Substance Abuse and Mental Health Services Administration \cite{samhsas_2014}, are critical for designing platforms where survivors feel safe to reconstruct their identities online \cite{Scott2023, randazzo2023}. Despite the importance of these principles, scholars have yet to \textit{model} the specific pathways or mechanisms by which survivors reconstruct their identities in online communities.
To address this gap, we drew on 30 in-depth qualitative interviews with trauma survivors and proposed a new \textit{Sociotechnical Model of Identity Integration}. Our model illustrates how online 
identity reconstruction unfolds in ways that can lead survivors toward an integrated identity \cite{herman1992}, or , alternatively, intensify fractures \cite{janoff-bulman_shattered_2002}.  Through this research, we extend Janoff-Bulman’s \cite{janoff-bulman_shattered_2002} concept of `the shattered self' by situating these fractures on identity frames \cite{hecht1993}, showing how online platforms, community norms, and algorithms can either widen and exacerbate or glaze and renew fractures.

In this paper, we use the art of Kintsugi as a guiding metaphor for illustrating when and how fractures can become integrative “golden seams” in a survivor’s identity. Like Kintsugi, successful identity repair blends the trauma (“cracks”) into a cohesive sense of self. Importantly, our findings reveal that Kintsugi represents a possible positive pathway in our sociotechnical model. If platform design, community support, and individual agency are well aligned, survivors can glaze these fractures and achieve identity integration. Conversely, misaligned design decisions—such as algorithmic grief bubbles or restrictive moderation—can widen fractures, akin to a failed Kintsugi repair. In doing so, we challenge Goffman’s \cite{goffman_presentation_1956} assumption of identity salience, arguing that improvisation often supersedes scripted presentations. Additionally, we identify \textit{grief bubbles} (algorithmic echo chambers that reinforce the traumatic identity) as a key barrier to identity integration. We also demonstrate how identity gaps between personal-relational or personal-communal frames spur survivors to \textit{seek out} supportive online spaces. Finally, we provide design recommendations that collectively embody trauma-informed design principles \cite{chen_trauma-informed_2022, Scott2023}.

Reflecting on Hemingway’s words above, we posit that being “strong at the broken pieces” can be facilitated through trauma-informed online communities. In the sections that follow, we present qualitative evidence and analysis that illuminate these sociotechnical pathways to identity integration, illustrating how the Kintsugi-based approach can transform survivors’ fractured identities into a renewed sense of self.

\section{PRIOR WORK}
We begin with a definition of trauma and an overview of perspectives on identity reconstruction following traumatic events. In the following section, we discuss how identity gaps can drive users to online communities for identity reconstruction and potentially emerge as a result of the design and structure of these environments. Then, we detail the significance and application of trauma-informed design principles. Collectively, this literature provides a comprehensive foundation for understanding the multifaceted nature of trauma, the dynamics of identity reconstruction in online settings, and the role of trauma-informed principles in investigating the online experiences of trauma survivors.

\subsection{Identity Reconstruction After Trauma}
In the following section, we delve into the complexities of trauma and identity reconstruction by exploring how traumatic experiences can shatter one’s sense of self, necessitating the communicative process of identity reconstruction. We then draw from Hecht’s \cite{hecht1993, kuiper2023bridging} communication theory of identity (CTI) to examine how identities are co-constructed through and within communication across personal, communal, relational, and enacted frames.
\label{sec:trauma}

\subsubsection{Defining trauma} Trauma, an emotional reaction to an unsettling event \cite{samhsas_2014}, can have ``a long-lasting effect on the self and psyche’’ \cite[P.4]{shapiro2010trauma}. Giller \cite{giller1999psychological} argues that what qualifies as traumatic is dependent on the perception of the individual. As van der Kolk and McFarlane \cite{van2014body} explain, the meaning that survivors attach to a traumatic experience is as important as the trauma itself, which speaks to the significance of personal interpretation. Two people can experience similar traumatic events yet interpret and internalize the trauma differently \cite{giller1999psychological}. To categorize trauma types, we pull from pre-established lists which include but are not limited to sexual assault, intimate partner violence (IPV), psychological maltreatment, community violence, and environmental disasters \cite{traumatypes2008}. Trauma is ubiquitous—prior work indicates that encountering multiple traumatic experiences is commonplace \cite{kessler2013posttraumatic, norris1992epidemiology}. Benjet et al. \cite{benjet_epidemiology_2016} estimate that 70\% of adults worldwide have undergone at least one traumatic experience in their lifetime. Chen et al. \cite{chen_trauma-informed_2022} argue that the majority of technology users are trauma survivors given the widespread nature of trauma. Considering trauma’s pervasiveness, it is imperative for research to gain an in-depth understanding of how survivors reconstruct their identities in online spaces. 

\subsubsection{Fractured identity: The shattered self} \label{sec:fractured_related} Traumatic experiences can shatter one’s sense of self \cite{janoff-bulman_shattered_2002}, permanently changing their identities  \cite{herman1992, schorer1990}. Reconstructing one’s identity after trauma is done ``in the presence of others’’ \cite[P.25]{egnew2005meaning}, meaning that redefining one’s self-concept manifests through communicative behaviors (e.g., talking, writing, listening; \cite{dundas_finding_2021, duma_womens_2007, dunn2010judging, pennebaker1988disclosure}. Experiencing what Egnew \cite [P.255]{egnew2005meaning} describes as ``wholeness as a person’’ has been found to alleviate the psychological effects of trauma like post-traumatic stress disorder (PTSD), anxiety, and depression \cite{brewin2000meta, kira2018trauma, kira2019toward, kira2017threatened, ozer2003predictors}. Thus, the process of identity reconstruction in reaching ``wholeness’’ can help protect users from psychological harm \cite{herman1992, szabo2015identity}). In resilience studies, achieving wholeness is described as `bouncing back’ to an original form \cite{buzzanell2010resilience}, which can be challenging for survivors due to the enduring nature of trauma. 

\begin{figure*}[ht]
  \centering
  \includegraphics[width=0.5\textwidth]{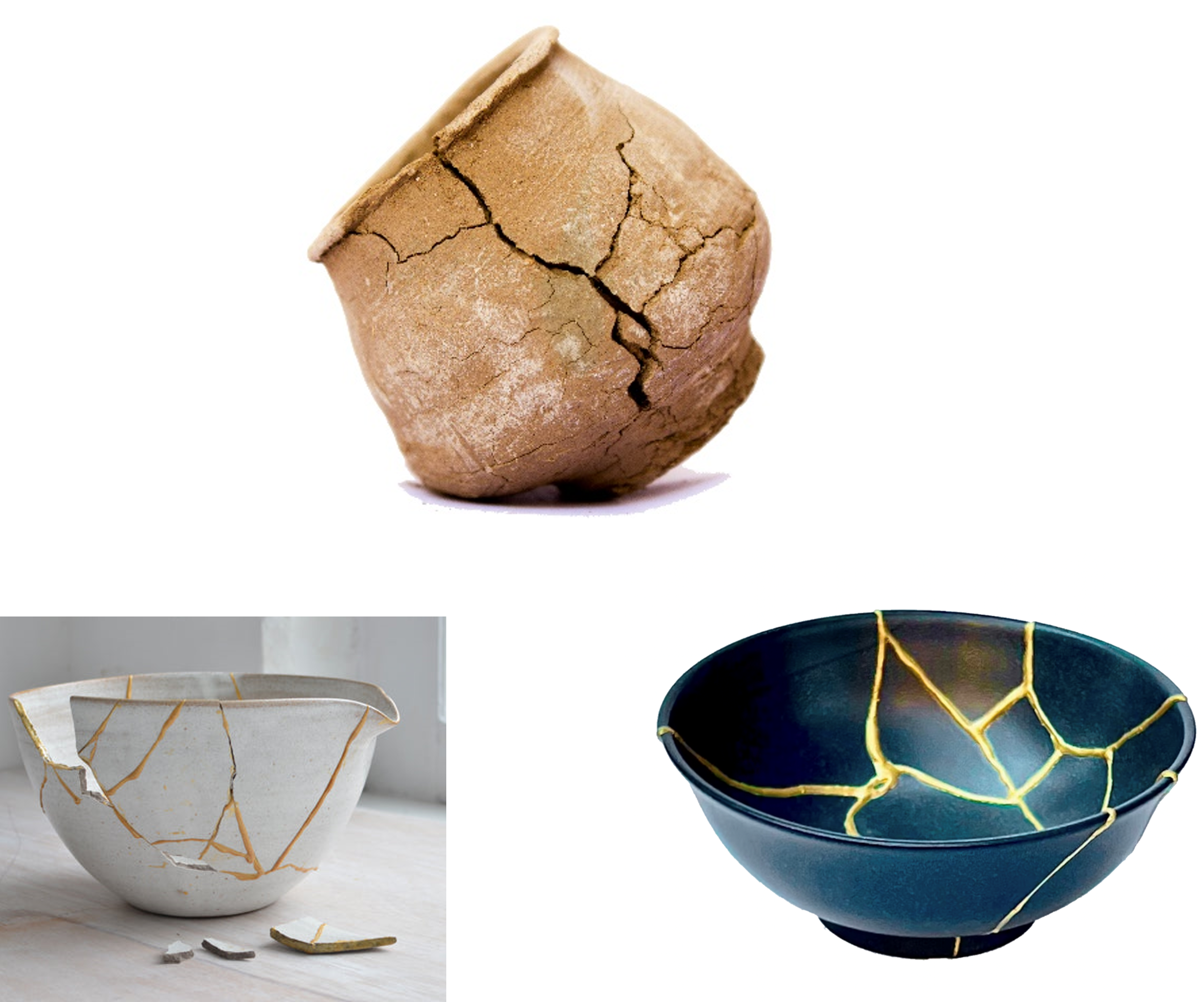}
  \caption{The Japanese art of Kintsugi where broken pottery is repaired by introducing gold into the cracks. We use this metaphor to represent the mending of fractured identities. The image on the top shows a cracked pot. The image on the bottom left shows a failed attempt at Kintsugi as described by a blogger who attempted the art form \cite{noauthor_kintsugi_nodate}. The image on the bottom right shows a successful attempt at Kintsugi where the cracks in the pottery are glazed.}
  \Description{This figure shows a piece of pottery that has undergone the art of Kintsugi, the Japanese tradition of repairing broken pottery with gold.}
  \label{fig:glazedpottery}
\end{figure*}

Herman \cite[P.241]{herman1992} argues that the ``resolution of the trauma is never final’’ and that achieving wholeness requires integrating the trauma into one’s identity (i.e., integrated identity; \cite{herman1992}). In this way, identity reconstruction transforms ``the traumatic memory so that it can be integrated into the survivor’s life story’’ \cite[P.202]{herman1992}. Herman’s \cite{herman1992} theorizing more closely resonates with Kintsugi, the Japanese art of mending broken pottery \cite{koren1994wabi}.\footnote{In English, Kintsugi translates to golden journey \cite{koren1994wabi}.} Instead of throwing a broken vase away, the Kintsugi method glues the broken pieces together using golden paint. This process not only improves the aesthetic but honors the cracks as part of the object’s history \cite{koren1994wabi} as seen in Figure \ref{fig:glazedpottery}. The Kintsugi art form can be used as a metaphor for Herman’s \cite{herman1992} integrated identity as the broken vase is seen as more beautiful for its imperfections. Despite these theories, work is limited in how identity reconstruction unfolds \textit{online} for trauma survivors \cite{dym2019coming} and how the design of digital spaces influences the path toward identity integration \cite{herman1992}.

\subsubsection{Frames of identity} To better understand the communicative process of identity reconstruction, we pull from Hecht’s \cite{hecht1993} communication theory of identity (CTI) which helps us assess how identity changes and evolves \cite{Littlejohn2009encyclopedia} through and within communication. CTI argues that individuals have multiple frames of identity which can be overlapping and at times contradictory \cite{hecht1993}. These frames include the\textbf{enacted}, \textbf{personal}, \textbf{communal}, and \textbf{relational} \cite{hecht1993}.  

The \textit{enacted frame} of identity is how we express ourselves through behavior, echoing Goffman’s presentation of self. The vast majority of HCI and computer-mediated communication (CMC) studies on identity use Goffman's \cite{goffman_presentation_1956} dramaturgical framework \cite{huang2021literature}. In this framework, Goffman \cite{goffman_presentation_1956} argues that individuals enact roles in a planned performance for the purpose of impression management.\footnote{Marwick and boyd \cite{marwick2011}, for example, introduce the concept of context collapse, where various audiences—akin to Goffman's \cite{goffman_presentation_1956} societal stages—merge due to the affordances of social media} Goffman \cite[P.21]{goffman_presentation_1956} posits that individuals meticulously design their social scripts, "with painstaking care, testing one phrase after another, in order to follow the content, language, rhythm, and pace of everyday talk."\footnote{Social scripts are constructed guidelines that people follow in particular social situations or roles \cite{improvedefine}.} While Goffman's work \cite{goffman_presentation_1956} is foundational, this study acknowledges the multifaceted nature of identities. In other words, identity extends beyond mere performance; it is also co-constructed through communication with oneself and others.

\begin{figure*}[ht]
  \centering
 \includegraphics[width=0.7\textwidth]{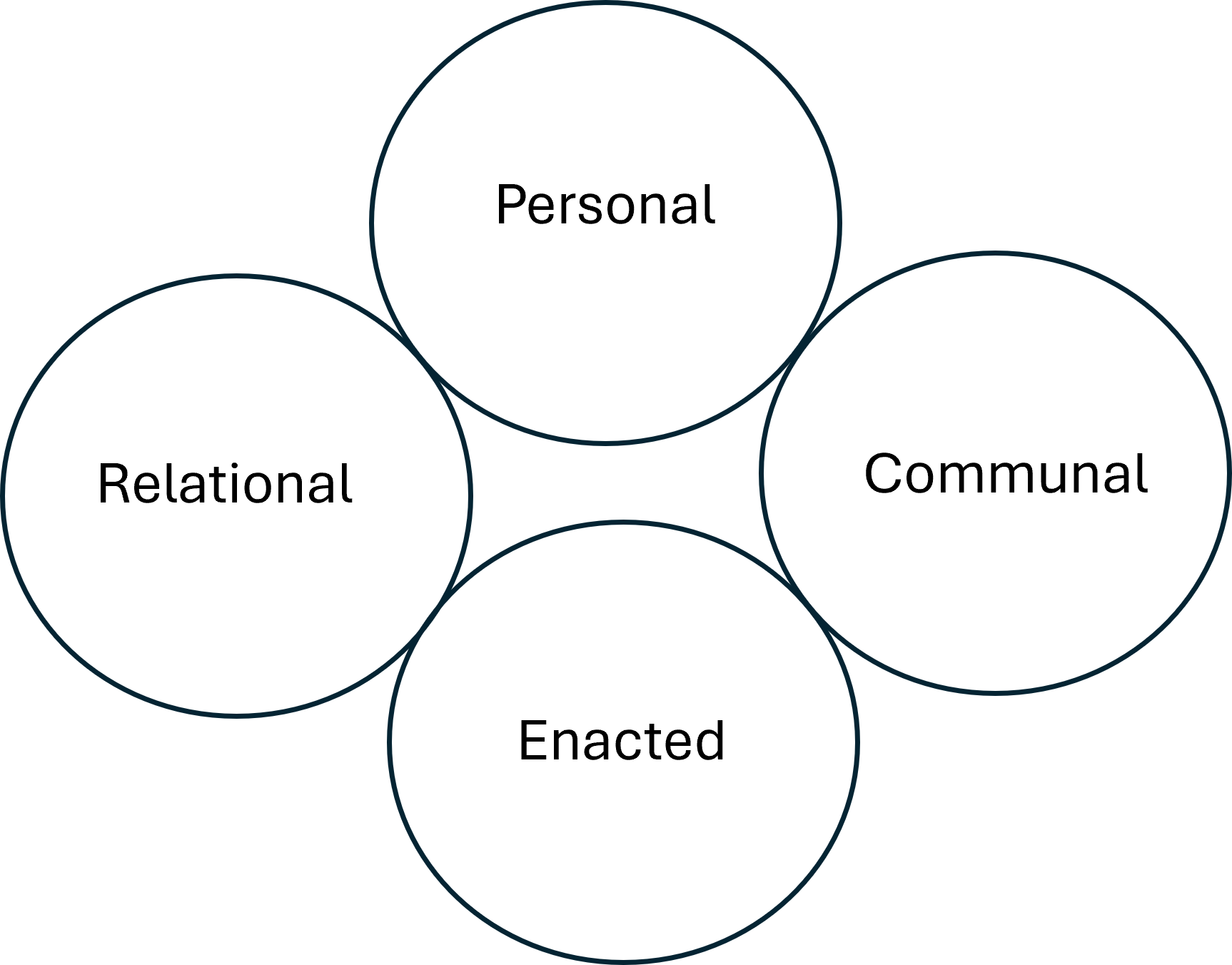}
  \caption{The figure, adapted from \cite{kuiper2023bridging}, visualizes Hecht's \cite{hecht1993} communication theory of identity}
  \Description{This figure is a replication of Hecht's (1991) visual of the CTI frames from Kupier (2023).}
  \label{fig:CTI1}
\end{figure*}

The \textit{personal frame} describes an individual’s self-concept or how we see ourselves. The \textit{communal frame} is how we view ourselves as part communities or groups (e.g., employees at the same company, members of a congregation). The \textit{relational frame} reflects who we are to others, which can be co-created (e.g., spouse, brother, friend) or ascribed (i.e., labels others assign to us) \cite{hecht1993}.

\subsection{Navigating Identity Gaps in Online Communities}
In this section, we delve into the concept of identity gaps (i.e., contradictions among identity frames; \cite{jung2004elaborating}), which we explore as a key factor in online identity reconstruction (i.e., redefining one’s self concept in online spaces; \cite{huang2021literature}).\footnote{Identity reconstruction often unfolds in the immediacy of face-to-face interactions \cite{bessiere2007ideal, toma2010looks}.} We examine how individuals redefine their self-concept in online communities and how identity gaps that potentially emerge from within groups can impact the online experiences of users. \label{sec:frames}

Before delving into identity gaps, we must first define online communities which we focus on due to the pivotal role they can play in online identity reconstruction {\cite{Andalibigarcia2021, ammari_crafting_2017, andalibi_announcing_2018, haimson_trans_2020, Haimsonoli2018, Semaanetal2016, Semaanetal2017, craig2014you, ammari_accessing_2014, ammari_thanks_2016, ammari_networked_2015, Blackwell2016}}. Drawing from Lyons et al. \cite{lyons2012towards}, we define online communities as groups hosted on social media platforms (i.e., dynamic technologies designed to enable the creation and exchange of content, ideas, interests, and various expressions within online communities or networks \cite{lee2003virtual, obar2015social, kietzmann2011social}). Users in these communities are often driven by a desire to interact with others \cite{lin2007role} around shared needs or interests \cite{ganley2009ties, preece_top_2004}. While our primary focus is on the dynamics of online communities, we also explore the influence of social media platforms on these communities through their design and governance structures \cite{gillespie2018custodians}. These online groups can range from tightly-knit communities like Subreddits and Facebook Groups \cite{gibbs2022digital} to more loosely connected networks bound by hashtags and shared discourses such as on Twitter \cite{jackson2020hashtagactivism, shugars2019keep} or thematic clusters curated algorithmically, often found on TikTok \cite{barta_constructing_2021, DeVito2021, randazzo2023}.

\subsubsection{Personal-relational identity gaps} Prior work has found that individuals reconstruct their identities online to avoid stigmatization (i.e., disapproval of someone based on their identity; \cite{andalibi_what_2019, andalibi2016, ammari_self-declared_2019}. Stigmatization from one’s friends or family can lead to the formation of identity gaps (i.e., contradictions among identity frames; \cite{jung2004elaborating}), driving users to reconstruct their identities online. Personal-relational gaps, for example, materialize when there are discrepancies between how individuals see themselves and the labels others ascribe to them \cite{jung2004elaborating}. In the context of sexual assault, survivors are often labeled `passive victims’ or `blameworthy victims’ \cite{dumont2003social, sudderth1998ll}. The passive victim is unfairly described as weak, susceptible, and always dependent on others for help \cite{dundas_finding_2021, dunn2010judging, mckenzie2011telling}. In contrast, the label of blameworthy victim views survivors as responsible for the assault \cite{dumont2003social,sudderth1998ll}, risking retraumatization (i.e., stress reactions that feel similar to the original traumatic event; \cite{duckworth2012retraumatization}). Greeson et al. \cite{greeson_nobody_2016}, for instance, found that a negative disclosure experience with law enforcement can leave survivors feeling as if they are to blame for sexual assault, resulting in secondary victimization (i.e., victim-blaming practices of service providers that result in additional trauma; \cite{campbell2005sexual}), causing survivors to lose faith in social systems \cite{greeson_nobody_2016}. These ascribed labels can differ from how survivors see themselves and lead to personal-relational gaps that interfere with the ``ability to receive support and comfort’’ \cite[P.269]{ahrens_being_2006}, posing risks for identity integration \cite{dunn2010judging}.

\subsubsection{Personal-communal gaps}Traumatic events ``shatter the sense of connection between individual and community’’ \cite[P.71]{herman1992} which can be due to personal-communal gaps. This type of gap occurs when there are differences between someone’s self-concept and how they are perceived by or relate to their communities \cite{jung2004elaborating}. Semaan’s \cite{Semaanetal2016, Semaanetal2017} work, while not explicitly using identity gaps, still provides evidence of personal-communal gaps arising for veterans (a population at-risk of experiencing trauma \cite{veteranaffairs2023}) transitioning back to civilian life. In these studies, their personal identity is shaped by the military which includes experiences like combat, living in a structured environment, and being part of a close-knit unit. These situations contribute to a personal identity that is deeply rooted in one’s military experience. When veterans return home, their personal identity can contradict the identity of their communities, especially if friends or family struggle to understand the psychological impact of war  \cite{Semaanetal2016, Semaanetal2017}. As Semaan’s  \cite{Semaanetal2016, Semaanetal2017} research demonstrates how the use of online communities, hosted on social media platforms, helps ease the return to civilian life. Such spaces guide veterans in reconstructing new identities and connecting with peers, often in online spaces separate from their primary social circles \cite{Semaanetal2017}.

\subsubsection{Personal-enacted gaps} Research using CTI is limited in providing evidence of how online identity reconstruction can lead to identity gaps. Drawing from prior work on sociomateriality \cite{barad2007meeting, suchman2008feminist, orlikowski200810, baym2010making}, which we discuss in section \ref{sec:trauma}, we suspect that personal-enacted identity gaps, ``the difference between an individual’s self-view and the self that one expresses in communication'' \cite{amado2020mind} can influence how users understand and present their identities in digital spaces \cite{devito_platforms_2017}. Algorithms, for example, can play a role in how users understand or perceive their identities \cite{DeVito2018, DeVito2021}. Trauma survivors in Randazzo and Ammari \cite[P.5]{randazzo2023} refer to content recommendation algorithms as  ``perfect little mirrors’’ due to their ability to collect context-aware data from users and personalize online experiences to reflect one’s preferences \cite[P.72]{lee_what_2010}. These types of algorithms can lead to filter bubbles which are curated news feeds that filter out contradicting information \cite{pariser_filter_2011}. Machine learning algorithms frequently group users into identity categories and make content recommendations based on inferences from their digital traces \cite{kang2020feeling}. Soffer \cite{soffer2021algorithmic} refers to these user groups as ``liquid social categories’’ that often change due to an algorithm’s interpretation of a user’s digital trace data. Using data in real-time results in ``a shift in the algorithmic output and its identification with a certain category’’ \cite[P.308]{soffer2021algorithmic}. Gillespie \cite[P.183]{gillespie2014relevance} describes this process as an ``entanglement. . . [of] recursive loops between the calculations of the algorithm and the calculations of people.’’ These experiences can lead to algorithmic anxiety (i.e., uncertainties about how algorithms operate; \cite{Jhaveretal2018}) and algorithmic symbolic annihilation (i.e., algorithms inadvertently supporting common and oversimplified narratives \cite{Andalibigarcia2021}. Collectively, this work helps us to provide empirical evidence for personal-enacted gaps in the context of HCI and emerge due to the design of online communities. 

\subsection{Trauma-Informed Design}
In this section, we examine trauma-informed design frameworks \cite{chen_trauma-informed_2022, Scott2023} as they apply to the design of online communities. The principles of these frameworks help contextualize how affordances, features, and governance structures can support survivors' identity reconstruction in online communities. \label{sec:traumadesign}

The trauma-informed design framework focuses on features (e.g., algorithms, comment sorting), governance (e.g., moderators, flagging; \cite{Jhaveretal2018}), and technological affordances (i.e., opportunities for action; \cite{parchoma_contested_2014}) that are prevalent in online communities hosted on social media platforms. In this paper, we draw from Chen et al.'s \cite{chen_trauma-informed_2022} trauma-informed computing and Scott et al.'s \cite{Scott2023} social media design frameworks which adapt trauma-informed principles established by the Substance Abuse and Mental Health Services Administration (SAMHSA) to communication technologies. These principles include (a) safety (i.e., sense of security while using a platform; \cite{ammari_moderation_2022,Jhaveretal2018, takahashi_potential_2009}; (b) trustworthiness and transparency (i.e., features that help build user trust; \cite{Bellini2023, dym_social_2020}); (c) peer support (i.e., elements that facilitate shared experiences and emotional support; \cite{andalibi2016, andalibi_announcing_2018, blanch_social_2012, randazzo2023}); and (d) empowerment, voice, and choice (i.e., features that enhance user agency; \cite{andalibi_announcing_2018, haimson_trans_2020, Im2021, randazzo2023}). Both frameworks \cite{Scott2023, chen_trauma-informed_2022} are largely theoretical at this stage, creating an opportunity for more empirical work to substantiate their claims. Therefore, we situate prior HCI studies in the framework of these principles to identify potential areas of concern for survivors. 

\subsubsection{Supportive online spaces} In trauma-informed design, peer support refers to the facilitation of connections between users who share similar experiences, offering them a space for mutual assistance and understanding \cite{chen_trauma-informed_2022, Scott2023}. Online social support is well-studied in HCI \cite{pearce2022online, liu2018meta, ellison2022scholarship}, particularly in relation to life transitions. Scholars in this space argue that online communities are more than passive tools but are active spaces where users can renegotiate and strengthen their identities during pivotal life changes \cite{Andalibigarcia2021, ammari_crafting_2017, andalibi_announcing_2018, haimson_trans_2020}, referring to this phenomenon as social transition machinery \cite{Haimsonoli2018}. Such changes include, but are not limited to, users returning from military service \cite{Semaanetal2017, Semaanetal2016}, coming out as LGBTQIA+ \cite{dym2019coming, craig2014you}, becoming new parents \cite{ammari_accessing_2014, ammari_thanks_2016, ammari_crafting_2017} or being a parent to children with special needs \cite{ammari_networked_2015, ammari_accessing_2014}or LGBTQIA+ youth \cite{Blackwell2016}. While transitional periods can be challenging or involve trauma, these experiences are not inherently traumatic. In contrast, trauma involves unexpected and severe events that can threaten one’s sense of safety and well-being \cite{van2014body}. Whereas transitions, more often than not, offer a path for gradual change and adjustment. Trauma, on the other hand, forces a sudden and non-linear path in how one perceives oneself and the world. Still, studies on life transitions can help contextualize supportive elements such as validation, disclosing experiences, and exchanging coping strategies \cite{de2014mental, haimson_disclosure_2015, bazarova_self-disclosure_2014, pearce2022online}.

Supportive communities can also take the form of alternative safe spaces \cite{randazzo2023} which have been found to bear the weight of identity-based stigmas \cite{dym2019coming}, allowing trauma survivors to `tip-toe' up to trauma narratives from a safe distance \cite{ammari_crafting_2017, dym2019coming}. Safe spaces can manifest as Facebook Groups, subreddits, or TikTok communities and often rely on a platform's transportability\footnote{The extent
to which platforms afford transportation to new communities or worlds of information; \cite{randazzo2023}.} Peer support often involves receiving direct \cite{devito_platforms_2017} or indirect feedback \cite{randazzo2023}. Both types of feedback are similar in that they both involve likes, comments, or replies. The difference, however, is in how communication is directed-either \textit{directly} to the publisher of the post \cite{DeVito2018} or \textit{indirectly} to a user internalizing the comments despite not being post's author \cite{randazzo2023}. As Randazzo and Ammari \cite{randazzo2023} explain, indirect feedback can offer survivors an opportunity to gauge the potential response they might receive if they were to disclose their own experiences, thereby serving as a barometer for peer support. Given that earlier studies establish the importance of social media for coping with change, we ask,
\begin{quote}
  \textbf{RQ1: What is the function of online communities in the process of identity reconstruction for trauma survivors?}  
\end{quote}

\subsubsection{User safety} In online communities, effective moderation is paramount to user safety \cite{ammari_moderation_2022, Jhaveretal2018, jhaver2017view}. Volunteer moderators are often praised for maintaining a platform's integrity \cite{gillespie2018custodians, Jhaveretal2018} while navigating complex landscapes of interlocking identities and power structures (e.g., women in patriarchal societies; \cite{ammari_moderation_2022}). Tools that allow for trigger warnings can help moderators ensure user safety \cite{chen_trauma-informed_2022, Scott2023, andalibi_announcing_2018, haimson_trans_2020, jhaver_designing_2022, Im2021, randazzo2023}. The function of trigger warnings is twofold. First, they can act as a preemptive safeguard, notifying users about potentially distressing content, providing them with the agency to make an informed decision about their engagement \cite{branley2017pro}. Second, trigger warnings can signal a community's commitment to the emotional and psychological well-being of its users, thereby enhancing trustworthiness \cite{haimson_trans_2020, jhaver_designing_2022}. Yet the efficacy and appropriateness of trigger warnings are still subject to debate. Boysen \cite{boysen2017evidence} argues that such warnings do not effectively mitigate psychological harm and can even foster behaviors (e.g., avoidance) that could be counter to trauma recovery. Still, the role of trigger warnings in online communities, particularly in environments where sensitive topics are discussed, underlines the importance of acknowledging the diverse experiences and needs of users. As such, trigger warnings remain a critical area of focus for trauma-informed design strategies.

\subsubsection{Transparency and trustworthiness} Transparency about community decisions is relevant to trauma survivors \cite{randazzo2023}. Randazzo and Ammari \cite{randazzo2023}, for example, find that RedditCares is being used as a form of harassment instead of as a mental health resource. RedditCares uses a community-flagging mechanism to identify users that appear to be undergoing mental health distress. The authors \cite{randazzo2023} suggest the need for platforms to be transparent about the tool's efficacy. In addition to recruiting community members, platforms can also incorporate machine-learning algorithms to flag harmful content \cite{gillespie2018custodians, jhaver2019human}. Gillespie \cite[P.98]{gillespie2018custodians} argues that the task is complex ``given that offense depends so critically on both interpretation and context.’’ Algorithms can face difficulty discerning which content is harmful, particularly due to the `stickiness' of emotional content, leading algorithms to prioritize affective media in users' feeds \cite{tufekci2015algorithmic, kramer2014experimental}. Matias \cite{matias2019preventing}, for example, found that emotional content that ranges from uplifting to distressing has been shown to drive higher user engagement metrics (e.g., session durations). These algorithmic choices have implications for users' emotional well-being. Constant exposure to certain types of emotional content, particularly negative or distressing material, has been found to affect users' emotional states \cite{Jhaveretal2018, davidson2020prejudice}. The prioritization of emotional content serves as a double-edged sword as it can bolster user engagement for the platform while also undermining users' well-being, raising questions regarding the ethical implications of algorithmic decision-making \cite{bucher2019algorithmic,  vaidhyanathan2018antisocial}. Ensuring transparency, however, requires a balanced approach that combines both algorithmic detection and human oversight \cite{jhaver2019human}. Algorithms, while not an ideal solution on their own, can aid in identifying harmful content, particularly when combined with human intervention \cite{ammari_moderation_2022}. As a result, automated detection algorithms have the potential to contribute to creating a supportive environment for identity reconstruction after trauma \cite{andalibi2016, randazzo2023}. Specifically, given the needs of traumatized individuals for support from different sources, algorithmic recommendations act like a diamond that allow users to experience ``diffracted connections with groups'' discussing similar topics \cite{lee2022algorithmic}[p.11] to find what they need.  However, the current state of algorithmic recommendation and moderation can cause psychological distress over privacy, safety, place in society, and identity (re)construction \cite{Karizat_et_al_21} which in turn keeps at-risk populations from crucial information because they do not want to engage in content that they ``feel they cannot escape.'' \cite{milton_et_al_23}

Given the difficulty of designing online communities that afford users better support, and the challenging nature of moderation for traumatized users, we ask,

\begin{quote}
    \textbf{RQ2: How can trauma-informed design recommendations better support online identity reconstruction after trauma?}
\end{quote}

\section{STUDY DESIGN}

This study adopts a qualitative research design which ``is appropriate when the purpose of the research is to unravel complicated relationships'' \cite[P.51]{rubin_qualitative_1995}. Specifically, we leveraged semi-structured interviews to gain an in-depth look at the function of digital platforms in the identity reconstruction process for users post-trauma. We then draw on 30 interviews, collected after receiving Institutional Review Board (IRB) approval, as part of a larger project on the experiences of survivors online.

\subsection{Data Collection}

We recruited participants from 70 online groups focused on trauma recovery (e.g., domestic violence survivors, rape survivors) and from alternative safe spaces (i.e., trauma-adjacent groups; \cite{randazzo2023}) focused on true crime media (e.g., podcasts, documentaries) on Reddit and Facebook. The authors searched for these communities using keywords developed after reviewing earlier work (e.g., trauma, traumatic experiences, social support). Additionally, we disseminated our recruitment messages on social media platforms like Twitter and TikTok. However, we noticed lower traction using the latter recruitment platforms. Earlier work demonstrates a connection between true crime and trauma survivors \cite{boling_fundamentally_2020, boling_undisclosed_2018, pavelko_muderinos_2020, myles_not_2018}. Boling \cite{boling_fundamentally_2020}, for example, found that IPV survivors describe true crime communities as spaces where similarly traumatizing experiences are discussed openly, allowing survivors to produce counter-narratives and challenge societal norms \cite{boling2018undisclosed}. Additionally, Pavelko and Gall-Myrick \cite{pavelko_muderinos_2020} found that multiplatform exposure (i.e., engaging in online communities for true crime podcasts on multiple platforms) improves the well-being of individuals diagnosed with a mental illness. For these reasons, we recruited from alternative (e.g., true crime communities) and traditional (e.g., trauma support groups) for this study. 

We began by asking administrators and moderators for permission to post a brief Qualtrics screening questionnaire that asked potential participants for demographic information, social media usage, and their informed-consent. We also asked respondents to describe their trauma broadly. To help them do so, we provided a number of suggested trauma categories as defined by The National Traumatic Stress Network \cite{traumatypes2008}. Because we understand that trauma can be complex, we also allowed respondents to choose "other" category and describe their trauma in a way that better reflected their experiences. Some of the responses under this category included: ``Long-term depression and severe anxiety,'' ``Parental abandonment,'' witnessing suicide attempts of family members of acquaintances, among others. 

We received 1,614 questionnaire responses and filtered out individuals that did not identify as a trauma survivor, reside in the United States, at least 18 years of age, and were part of online communities that discussed trauma narratives. Using purposive sampling, we asked 174 respondents to register for an interview and gave participants a \$20 gift card to Amazon as a thank you for their participation. We continued to conduct interviews until we reached data saturation and started hearing similar themes from our participants. This study draws on 30 interviews, a subset of a larger research project investigating the experiences of trauma survivors online. Interviewee details are presented in Tables 2 and 3.

All interviews were conducted on Zoom, and lasted, on average, 58 minutes (avg. 26-116 minutes), amounting to 55 total hours. We developed trauma-informed interview questions \cite{chen_trauma-informed_2022} that asked participants about their experiences using social media platforms, however, we did not stop participants from sharing details about their trauma to prevent participants from feeling silenced which can also be retraumatizing. Instead, we encouraged autonomy by leaving the decision to share details up to the participant. At the start of the interview, we reminded participants that (a) this is a safe space to discuss their trauma, however, sharing details is not required to participate; (b) none of the interview questions will delve into specifics about their trauma; and (c) we can skip any question that they do not want to answer. Additionally, we had resources available (e.g., support hotlines) if needed. Interviews began with questions about how participants discuss their trauma whether online or offline. We then asked the interviewees to describe their online experiences discussing trauma in their different online communities and on various platforms. We probed  interviewees to share both positive and negative experiences in their online interactions. We then asked about how they presented their identities in different online communities. Finally, we asked interviewees about how they are using social media in their current stage of recovery (at the time of the study) and asked them to reflect on their online interactions during past stages of recovery. 

\subsection{Participant Demographics}
The participants of this study were managing a diverse range of trauma experiences as seen in Tables \ref{tab:dem1} and \ref{tab:dem2}. These tables detail the subset of interviews we draw from for this study. The below demographics reflect the participants as part of the larger study. 78.9\% of the participants identified as female, 17.5\% as male, and 3.5\% as non-binary. Our race demographics include 1\% Asian or Asian American alone, 2\% Native Indian/American alone, 11\% Black or African American alone, 13\% Two or more races (Black/African American and Hispanic/Latino; Native Indian/Native American and White; Hispanic/Latino and White; Black/African American and White), 21\% Latino or Hispanic alone, and 52\% White alone. Participants represented 22 states across the United States. Educational backgrounds varied, with 1 participant reporting a high school diploma, 1 technical or vocational training, 11 reporting some college credit, 7 reporting an Associate's degree, 25 reporting a Bachelor's degree, 8 reporting a Master's degree, and 3 reporting a Doctorate/Professional degree as the highest level of completed education. 9\% of the participants reported their ages as 18-24, 37\% as 25-34, 35\% as 35-44, 16\% as 45-54, and 4\% as 55-64.

\begin{table*}[]
\caption{We split the table to accommodate space and formatting. See Tables \ref{tab:dem2} and \ref{tab:dem3} for a continued list of participants. (NB/TG: Non-binary/Third-gender; F: Female; M: Male)}
\label{tab:dem1}
\begin{tabular}{|l|p{280pt}|l|}
\hline
\multicolumn{1}{|c|}{\textbf{Pseudonym}} &
  \multicolumn{1}{c|}{\textbf{Trauma types}} &
  \multicolumn{1}{c|}{\textbf{Gender}} \\ \hline
Aurelia &
  Survivor of or witness to domestic violence, Emotional abuse or psychological maltreatment, Sexual abuse, Traumatic grief or separation, Parental or partner addiction or substance abuse &
  F \\ \hline
Betina &
  Emotional abuse or psychological maltreatment, Sexual abuse, Traumatic grief or separation, Parental or partner addiction or substance abuse &
  F \\ \hline
Bill &
  Emotional abuse or psychological maltreatment, Sexual abuse, Traumatic grief or separation, Other &
  M \\ \hline
Cameron &
  Survivor of or witness to domestic violence, Emotional abuse or psychological maltreatment, Sexual abuse, Traumatic grief or separation, Serious accident, illness, or medical procedure,Other &
  NB/TG \\ \hline
Caterina                        & Emotional abuse or psychological maltreatment, Sexual abuse, Physical abuse or assault, Other                                                                                                                                                                                                                                                                        & F                        \\ \hline
Courtney &
  Survivor of or witness to domestic violence, Emotional abuse or psychological maltreatment, Sexual abuse, Physical abuse or assault, Exposure to community violence, Traumatic grief or separation,Exposure to gun violence, Serious accident, illness, or medical procedure, Parental or partner addiction or substance abuse &
  F \\ \hline
Daisy                            & Survivor of or witness to domestic violence, Emotional abuse or psychological maltreatment, Sexual abuse, Physical abuse or assault, Traumatic grief or separation, Exposure to gun violence, Human trafficking,Parental or partner addiction or substance abuse                                                                             & F                        \\ \hline
Diane &
  Survivor of or witness to domestic violence &
  F \\ \hline
Dorothy &
  Survivor of or witness to domestic violence, Emotional abuse or psychological maltreatment, Sexual abuse,Physical abuse or assault, Parental or partner addiction or substance abuse &
  F \\ \hline
Eleanor                         & Emotional abuse or psychological maltreatment, Traumatic grief or separation,Other                                                                                                                                                                                                                                                                 & F                        \\ \hline
Eloise &
  Survivor of or witness to domestic violence, Emotional abuse or psychological maltreatment, Physical abuse or assault,Exposure to gun violence, Parental or partner addiction or substance abuse &
  F \\ \hline
Gabby                        & Survivor of or witness to domestic violence,Emotional abuse or psychological maltreatment, Sexual abuse, Traumatic grief or separation                                                                                                                                                                                                                                                                        & F                        \\ \hline
Grace &
  Emotional abuse or psychological maltreatment, Serious accident, illness, or medical procedure, Parental or partner addiction or substance abuse &
  F \\ \hline
Harmony &
  Traumatic grief or separation, Serious accident, illness, or medical procedure &
  F \\ \hline
Isabelle &
  Survivor of or witness to domestic violence,Emotional abuse or psychological maltreatment, Serious accident, illness, or medical procedure, Parental or partner addiction or substance abuse &
  F \\ \hline
\end{tabular}
\end{table*}

\subsection{Data Analysis}
Interviews were transcribed using Otter AI, a speech-to-text software, resulting in a corpus of 1,384 single-spaced pages. Any information that could potentially identify participants was substituted with pseudonyms. The lead author reviewed the first 12 transcripts, while four research assistants edited the rest of the transcriptions. All transcripts were subsequently imported into the NVivo software for more in-depth data analysis. We employed emergent theory as the analytical framework, which comprehends "why people behave or think as they do" \cite[P.269]{jaccard_theory_2020}. This approach emphasizes thorough description, a fresh understanding, and explanatory depth. Unlike comparable methodologies, emergent theory welcomes established theories and concepts (e.g., integrated identity, trauma-informed principles \cite{herman1992, Scott2023, chen_trauma-informed_2022}) during the data collection and interpretation phases, which allows for researchers to uncover new concepts while situating them in prior work. During the analysis, we conducted multiple rounds of constant comparison \cite{strauss_basics_1998} until overarching themes surfaced in our data. The authors met to discuss themes as they emerged. This allowed us to focus on important themes after each pass through the data. Broad segments were then further refined through open coding and axial coding to investigate the interconnections among the open codes \cite{corbin_basics_2008}. New themes emerging from the axial coding were also discussed at length between the authors utilizing a range of heuristic methods (see Jaccard \& Jacoby \cite{jaccard_theory_2020}) to analyze the data from various angles. True to the spirit of emergent theory, we continuously consulted existing literature throughout this process, thereby enabling relevant theoretical frameworks to naturally arise from the data and helping us tie segments to established concepts as well as develop new insights \cite{jaccard_theory_2020}. Some of the top themes included: (1) Identity integration and acceptance; (2) Personal-communal identity gaps; and (3) Personal-relational identity gaps. Some of the sub-themes associated with the top-level themes include: (a) Trauma Bonding; (b) Trauma Dumping; and (c) Trigger warnings, which we discuss in more detail in section \ref{sec:findings}.

\begin{table*}[h]
\caption{Continued list of interviewees}
\label{tab:dem2}
\begin{tabular}{|l|p{280pt}|l|}
\hline
\multicolumn{1}{|c|}{\textbf{Pseudonym}} & \multicolumn{1}{c|}{\textbf{Trauma types}}                                                                                                                                                                                                                                                                                  & \multicolumn{1}{c|}{\textbf{Gender}} \\ \hline
Katrina &
  Survivor of or witness to domestic violence, Emotional abuse or psychological maltreatment, Sexual abuse, Physical abuse or assault &
  F \\ \hline
Mila                            & Emotional abuse or psychological maltreatment, Sexual abuse, Physical abuse or assault, Exposure to community violence, Traumatic grief or separation                                                                                                                                                                                                                                              & F                        \\ \hline
Lars &
  Emotional abuse or psychological maltreatment, Sexual abuse,Physical abuse or assault, Traumatic grief or separation &
  M \\ \hline
Linda                           & Emotional abuse or psychological maltreatment, Parental or partner addiction or substance abuse,Other                                                                                                                                                                                                                                                                                                    & F                        \\ \hline
Luisa                           & Survivor of or witness to domestic violence, Emotional abuse or psychological maltreatment, Sexual abuse, Physical abuse or assault, Exposure to community violence, Traumatic grief or separation,Serious accident, illness, or medical procedure, Parental or partner addiction or substance abuse                                                                                                      & F                        \\ \hline
Lysa                          & Emotional abuse or psychological maltreatment, Sexual abuse, Physical abuse or assault, Traumatic grief or separation, Serious accident, illness, or medical procedure, Natural or man-made disasters, Military trauma,Other                                                                                                                                                                                                                                                                                  & F                       \\ \hline
Olive                           & Sexual abuse, Exposure to community violence, Exposure to gun violence,Serious accident, illness, or medical procedure, Parental or partner addiction or substance abuse                                                                                                                                                                                                                                 & NB/TG                    \\ \hline
Pat                             & Survivor of or witness to domestic violence, Emotional abuse or psychological maltreatment, Sexual abuse, Physical abuse or assault, Exposure to community violence, Traumatic grief or separation, Exposure to gun violence, Serious accident, illness, or medical procedure, Parental or partner addiction or substance abuse                                              & F                        \\ \hline
Rachel                        & Survivor of or witness to domestic violence, Emotional abuse or psychological maltreatment, Physical abuse or assault, Traumatic grief or separation                                                                                                                                                                                                                                                                        & F                        \\ \hline
Sharon                          & Emotional abuse or psychological maltreatment, Sexual abuse                                                                                                                                                                                                                                                                                                                                             & F                        \\ \hline
Stavros                         & Physical abuse or assault, Serious accident, illness, or medical procedure                                                                                                                                                                                                                                                                                                                              & M                        \\ \hline                   
\end{tabular}
\end{table*}

\begin{table*}[h]
\caption{Continued list of interviewees}
\label{tab:dem3}
\begin{tabular}{|l|p{280pt}|l|}
\hline
\multicolumn{1}{|c|}{\textbf{Pseudonym}} & \multicolumn{1}{c|}{\textbf{Trauma types}}                                                                                                                                                                                                                                                                                  & \multicolumn{1}{c|}{\textbf{Gender}} \\ \hline

Stella                          & Survivor of or witness to domestic violence, Sexual abuse, Physical abuse or assault, Exposure to gun violence, Serious accident, illness, or medical procedure, Parental or partner addiction or substance abuse                                                                                                                                                                                           & NB/TG                    \\ \hline
Stephanie                       & Survivor of or witness to domestic violence, Emotional abuse or psychological maltreatment, Physical abuse or assault, Traumatic grief or separation,Serious accident, illness, or medical procedure                                                                                                                                                                                                  & F                        \\ \hline
Tiffany &
  Survivor of or witness to domestic violence, Emotional abuse or psychological maltreatment, Sexual abuse, Physical abuse or assault, Other &
  F \\ \hline
Sally                            & Survivor of or witness to domestic violence, Emotional abuse or psychological maltreatment, Sexual abuse, Physical abuse or assault, Traumatic grief or separation, Exposure to gun violence, Serious accident, illness, or medical procedure, Parental or partner addiction or substance abuse                                                                                                                                                                                                                           & F                        \\ \hline
\end{tabular}
\end{table*}

\subsection{Researcher Positionality} 
The authors met weekly to engage in discussions about their positionalities throughout the entire research process. Both authors practiced reflexivity which required them to consistently acknowledge their potential biases to maintain the integrity of the project. The first author’s lived experiences deeply influence her approach to conducting research with trauma survivors. Her background informs the empathy and understanding she brings to each interview and shapes the dynamics of the interactions she has with participants. Despite sharing common ground, being a scholar and a white cis female can introduce a power dynamic that can potentially impact the openness of dialogue. This is where her positionality as a fellow survivor was helpful. Having shared experiences allowed her to engage empathically with participants, serving as a powerful conduit for building trust and rapport. With that said, the first author navigated these aspects of her identity carefully to ensure that interviews remained focused on participants’ narratives. 

The second author is a cis-man who was raised and educated in the Middle East.  His personal experiences growing up with physical and psychological trauma provided him with a point of reference to empathize with the experiences of participants. Drawing on his personal experience, the second author focuses his research on the use of technology in support of marginalized populations in different community contexts. 

In summary, the authors’ positionalities allowed them to connect deeply with participants. At the same time, they practiced reflexivity and self-awareness while navigating interviews. The authors remained cognizant of these dynamics throughout the project and strived to foster an environment where participants can share stories of trauma safely and comfortably.

\section{FINDINGS}
\label{sec:findings}

\begin{figure*}[ht]
  \centering
  \includegraphics[scale=0.5]{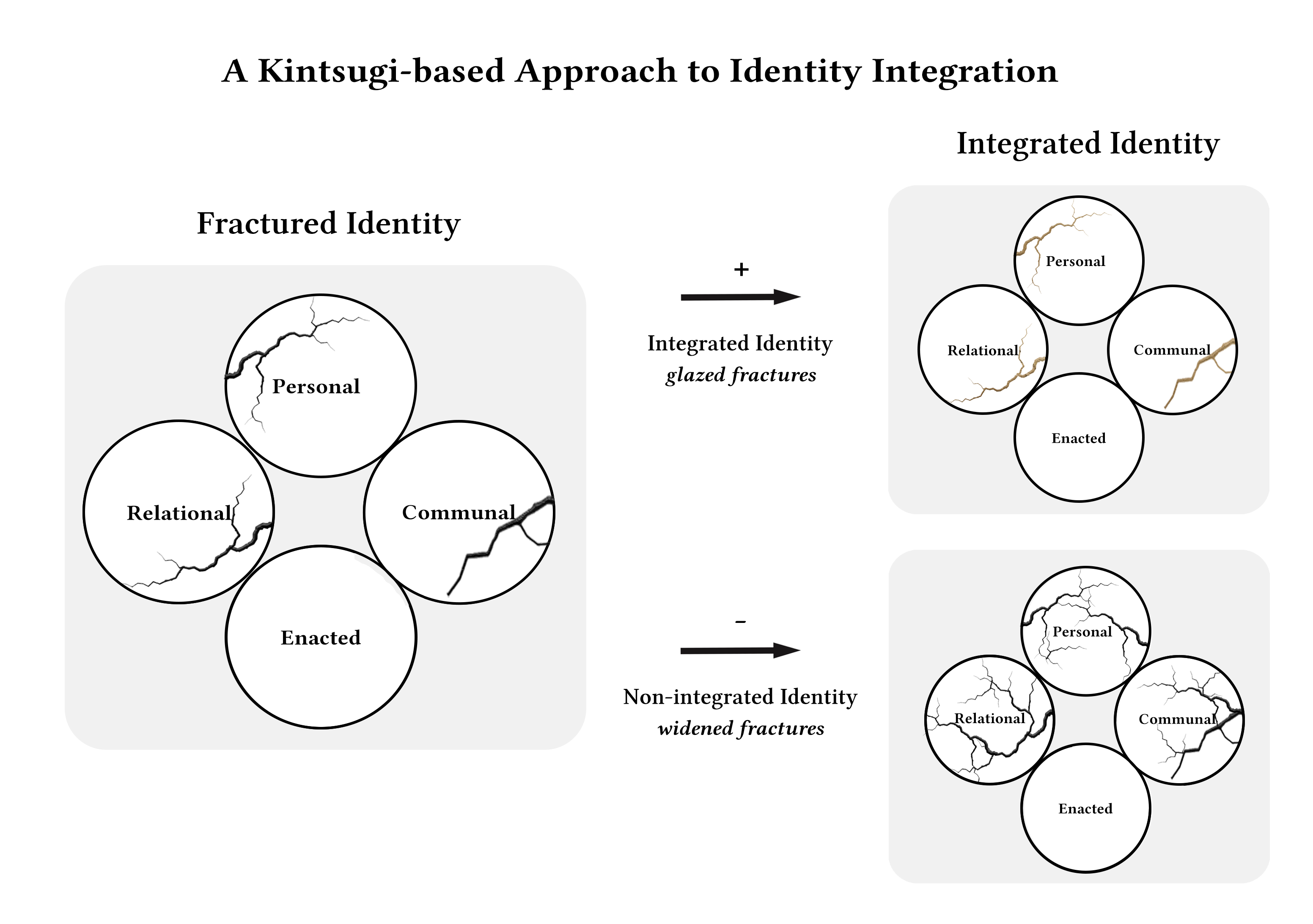}
  \caption{As is the case in Figure \ref{fig:glazedpottery}, this figure starts with a fractured identity as seen to the left. The bottom right shows a fractured identity where fractures were widened, leading to a non-integrated identity. The figure on the top right shows glazed fractures, and thus, an integrated identity.}
  \Description{This figure visualizes the identity fractures glazed in gold, reflecting acceptance of someone's traumatic experiences.}
  \label{fig:IdentityIntegration}
\end{figure*}

In our findings, we use Kintsugi as a metaphor to illustrate the process of reconstructing identities online after trauma. Much like identity reconstruction after trauma, Kintsugi demonstrates ``the propensity of repaired objects to embody dual perceptions of catastrophe and amelioration'' \cite{keulemans2016geo}. Our model (Fig. \ref{fig:IdentityIntegration}) begins with a fractured identity, presented in section \ref{sec:fractures}, as a result of the trauma. Fractures reflect the effects of trauma on an individual’s perception of self and manifest on identity frames (see \S\ref{sec:fractured_related}). Section \ref{sec:gapsonline} demonstrates how identity gaps can also result from online identity reconstruction (e.g., personal-communal, enacted-communal), thus allowing social media users to glaze their identity gaps. Section \ref{sec:gapsoffline} explains another side of this relationship by detailing how fractures can lead to identity gaps (e.g., personal-communal, personal-relational), thus increasing the user identity fractures.

Our findings reveal how trauma-informed design principles support identity integration (i.e., the process of weaving fractures into a new self-concept). In the following sections, we provide evidence for the different ways to weave fractures. In some, we show how engaging in online communities can support trauma survivors with fractured identities as they reconstruct their identities. For example, in \S\ref{sec:altspaces}, we describe how membership in a community not directly related to trauma can provide users with a space to reconstruct their fractured identities. However, we also show how misaligned design decisions become barriers for trauma survivors in online communities. For example, survivors can curate what we term \textit{grief bubbles} (\S\ref{sec:bubble}), echo chambers focused on discussions relating to their trauma, which might negatively affect their identity reconstruction.

\subsection{Fractured Identities}
\label{sec:fractures}
We observed a recurring theme across participants in our study: identity fractures. We find that these fractures are situated on identity frames and reflect the effects of trauma on one’s self concept, resulting in a fractured identity. As Stavros, a survivor of sexual assault, explains, ``I don't think anyone fully heals [from trauma]…it's like scar tissue that grows.’’ We visualize Stavros’s description of “scar tissue” as fractures that exist on identity frames. In this way, identity fractures become a permanent part of one’s self-concept and ``changes someone's life permanently,’’ noted by Tiffany, a survivor of assault. Stephanie’s experiences provide more evidence for the conceptualization of identity fractures. Stephanie is a former social worker who was arrested while advocating for the rights of parents wrongfully separated from their children. She began this advocacy work after observing Child Protective Services (CPS) disproportionately investigate and remove children from their homes in underserved communities. Stephanie was held in contempt for creating a video that sparked public demonstrations outside of a courtroom. With over a million views for one family's story, the online outcry created what Stephanie describes as ``a mess [and my] first time being on that side of things…arrested and locked up.’’ The charges against Stephanie were dropped after twenty-four hours in jail. Still, she was deeply affected by the experience, later being diagnosed with post-traumatic stress disorder (PTSD).

As Stephanie describes,``It really did something to me…have you ever done one day [in jail]? Have you ever done one hour, where your freedom was taken from you? The difference I have [experienced], is something [being] taken from you.’’ Here, Stephanie describes the impact of the fracture as ``something taken’’ from her understanding of self. Lars, a survivor of sexual abuse, expands on this idea by explaining that, ``Your identity can kind of dissolve a little bit when you're around an abusive person.’’ His trauma resulted in such extensive fractures that Lars found himself unable to recall who he was prior to the traumatic experience. As Lars said, ``It got to the point of depersonalization and derealization...I wasn't really able to rebuild whatever person I was before, because I kind of couldn't remember who that was.’’\footnote{Depersonalization is a psychological condition in which individuals feel detached from their own thoughts, feelings, or sense of self, as if they are observing themselves from outside their body. Derealization involves a similar detachment, but it is characterized by a feeling of disconnection from the surrounding world, making environments and objects seem unfamiliar or surreal \cite{american2013diagnostic}.} Francesca, a survivor of psychological maltreatment, resonates with Lars’s experience, sharing, ``I don’t know how to be myself because I lived so long in survival mode.’’ Not only do identity fractures have the potential to disrupt one’s self-concept, but they can also impair one’s ability to interact with others.

Courtney, along with other participants in the sample, describes not understanding ``how to \textit{human} appropriately’’ after trauma. Courtney is a survivor of parental addiction and a witness to intimate partner violence (IPV). She recalls her childhood as ``chaotic and adrenaline built.’’ Courtney recounts, ``My parents struggled a lot with addiction and violence and incarceration…When I was in eighth grade, my mom slit my stepdad’s throat and I had to spend all day cleaning up the blood.’’ These experiences left profound fractures on Courtney’s identity, persisting long after the traumatic events had passed. Similarly, Gabby, a survivor of sexual abuse, describes how it has taken her ``a long time to come to grasp with what happened’’ and she still tries ``to learn from the trauma every day,’’ which points to the enduring nature of identity fractures.
\subsection{Gaps Emerging From Online Identity Reconstruction}
\label{sec:gapsonline}
Our findings reveal a bidirectional relationship between identity gaps and online identity construction. In the below section, we detail how interactions in online communities can lead to personal-communal gaps as a result of community norms and group moderation. We also detail how personal-enacted gaps can emerge due to algorithmic systems.
\subsubsection{Trauma-aware grouping} Algorithms are integral to social media platforms and their online communities. Platforms often utilize group recommendation algorithms to guide users to relevant spaces. Our research indicates that these algorithms struggle to detect the subtleties of trauma, particularly in the context of IPV and sexual assault. As a result, Aurelia, along with other survivors in the sample, feels alienated as their fractured identities contradict with group norms in communities their algorithm recommended, leading to personal-communal gaps. Aurelia, for example, asked herself, ``Was it enough domestic abuse for me to really call myself a survivor? Or am I just claiming that name to sound like I'm part of the group?’’ In this quote, Aurelia questions the legitimacy of her experiences being ``enough’’ to claim survivorship. While Aurelia’s internal dialogue can be commonplace among IPV survivors, her thoughts are still deeply personal, making them difficult for an algorithm to detect. This experience prevents participants like Aurelia from engaging in online spaces due to the differing definitions of abuse and the risk of being disingenuous. Lars, a survivor of sexual assault, resonates with Aurelia’s challenge. When Lars experiences ``bad anxiety,’’ he heads to a support group on Reddit for men which was recommended by his system. Still, he finds little support as ``half the posts on there are more often [about] drugs. So, the trauma people are the minority there.’’ Reddit is affording low-transportability for Lars; his group recommendations are not transporting him to communities with peers that share similar traumatic experiences. In these examples, the algorithms are unable to recognize diverse experiences of trauma survivors, highlighting a sociotechnical gap which we explore in the discussion section. 

\subsubsection{The curation of grief bubbles} \label{sec:bubble} Sally, along with other survivors in the sample, describes ``hiding behind your trauma…as a way for people to get to know you but also not.’’ This behavior serves as a defense mechanism where individuals openly discuss their traumatic experiences to deflect from other aspects of their identities. Our findings suggest that ``hiding behind your trauma’’ can impede users from integrating their fractured identities into their self-concept. Stavros’s experience resonates with Sally’s quote and demonstrates how a lack of trauma-informed design can discourage identity integration. As a male survivor of assault, Stavros’s story became a commodity among advocacy and charitable organizations due to the usual reluctance of men to disclose their abuse. Stavros said, ``I was doing advocacy [work] in college and that was my entire identity, that just was who I was. And there was no script escaping that label.’’ At first, Stavros was not facing an identity gap but instead, experiencing identity fusion where his enacted role as an advocate eclipsed all other facets of his identity. Stavros's challenge was to detach his identity as a survivor from his overarching self-concept. Twitter supported his single-faceted identity by confining Stavros in what he refers to as a ``grief bubble’’ due to amplification and direct feedback, both technological affordances of Twitter's platform. Initially, Stavros thought that tweeting about the lack of protective policies for survivors was productive, but this perception changed with time. He shares, 
\begin{quote} 
``I thought anger was the most effective fuel—I ran on anger because you get positive reinforcement…[when] yelling into the void about, you know, `this policy is BS'...but it takes a lot out of you. You feel very drained at the end of it.’’
\end{quote}
As Stavros explains, enacting his survivor identity online afforded ``positive reinforcement’’ in the comments, retweets, and likes he received. Over time, Stavros realized that his online advocacy work was overshadowing other aspects of his identity; a personal-enacted gap emerged between Stavros and his content recommendations. It was not until Stavros left Twitter/X that he came to integrate his fractures as ``a piece of who I am [but] in a way that has become less important to my overall identity.’’ Stavros's path to integration points to the importance of empowerment, voice, and choice, one of the trauma-informed design principles. Had this principle been applied, Stavros would have been able to reconstruct his identity without being pulled further into a grief bubble. Pat,  whom we first introduce in section \ref{sec:gapsoffline}, also describes being caught in an algorithmically curated grief bubble. In the following excerpt, Pat reflects on how algorithmic systems are interfering with her online identity reconstruction,
\begin{quote}
 ``So, it's tailored to my interests. You're learning how I think...and you're not giving me a way to kind of get out of my echo chamber…I'm not seeing anything that makes me think differently and that's really frustrating for me because I'm somebody who likes to fight.’’
\end{quote}    
Pat describes herself as ``somebody who likes to fight’’ which suggests a preference for content that challenges her existing viewpoints. Yet, Pat perceives her algorithm to entrap her within an ``echo chamber’’ of content that reinforces the victim identity imposed by Pat's family, which we describe in section \ref{sec:gapsoffline}. This algorithmic behavior creates a personal-enacted gap, obstructing Pat's access to information that can help her integrate fractures into her identity. 

\subsubsection{Sorting for support} \label{sec:sorting} Online communities can allow users to sort comments based on popularity.\footnote{This feature is referred to as `Most Relevant' in Facebook Groups and `Top' within subreddits.} The aim of the sorting feature is to enhance user experience by prioritizing comments that are deemed most engaging, often based on the number of reactions and replies. We find evidence of this feature prioritizing feedback that is perceived as harmful by survivors in the sample, hindering their online identity reconstruction. For example, Daisy has to ``go through 20 or 30 comments…[to find] people who are truly sympathetic.’’ Daisy expects threads to be sorted in a way where she can receive indirect feedback. Instead, Daisy recounts that the comments are sorted in a way where ``it’s all people saying, `Well, that's not that bad’ and `Back in my day, we just had to grin and bear it.’’’ Daisy is hoping to come across supportive indirect feedback but instead is overwhelmed with insensitive comments that belittle her fractures. Rachel, a survivor of psychological maltreatment, explains how ``Certain generations try to minimize [trauma]…They'll be like, `You turned out fine’ or `I turned out fine.’ [But] what is fine? Define fine.’’ Rachel is describing how social media platforms often collapse contexts, meaning that online communities can include users of different generations and cultures with their own perspectives on healing from trauma. As Rachel notes, terms like ``fine’’ can have different meanings and interpretations. For survivors in the sample, exposure to language that is perceived as unsupportive can impede their online identity reconstruction. 

\subsubsection{Challenges in moderation}
\label{sec:mod}
Rachel, whom we first mention above in section \ref{sec:sorting}, describes herself as struggling ``with identity and self-love’’ due to being ``neglected as a child.’’ Reddit offers a feature called RedditCares that allows users to flag posts about suicide or self-harm. As Rachel explains, 
\begin{quote}
``When somebody says something that could be even remotely interpreted as a cry for help, somebody else on the sub can anonymously report them...But then people use it to kind of bully each other. So, somebody will be like, `I didn't mean I wanted to kill myself. Stop constantly reporting me.’’’ 
\end{quote}
In this quote, Rachel describes RedditCares as contributing to a personal-communal gap. This gap arises from a discrepancy between a user's personal needs and the communal dynamics of a subreddit which is influenced by the behavior of other users. Rachel also questions the effectiveness of this feature by asking, ``How useful is it…and whose responsibility is it to take care of other people on the internet?’’ Prior work points to group administrators and moderators taking on this responsibility. In the following paragraphs, we detail how Luisa and Stella manage this commitment in their roles as Facebook Group administrators. Luisa, a survivor of community violence, has a firm no-trauma-dumping rule which prevents members of a Facebook Group (dedicated to sharing information during the COVID-19 Pandemic) from disclosing their personal experiences. Luisa shares, \label{sec:triggerluisa} 
\begin{quote}
``Not allowing trauma dumping makes it a safer place for everyone...You can't control what anybody has been through. You have no idea what another person is carrying…And as much as we want to personalize things and make them more human, that could be rough for somebody going through comments.’’    
\end{quote}
In Luisa’s group, the no-trauma-dumping policy functions as a governance mechanism aimed at creating a safer environment for all members. Luisa recognizes that while disclosing trauma can make discussions ``more human,’’ these posts can also unintentionally retraumatize others. Still, restricting trauma disclosures limits the group's capacity to serve as an alternative safe space, where members can reconstruct their identities with others. One solution is to require trigger warnings on each post containing sensitive information which Stella discusses below. Stella, a survivor of verbal abuse, shares how they protect users from retraumatization while also allowing for sensitive discussions. 

Stella was part of a true crime podcast group whose members often discuss traumatizing events. Stella, understanding the impact of traumatizing content, asked the members to include trigger warnings at the top of their posts.\footnote{A trigger warning (TW) can look like: TW: self-harm, suicidal ideation.} Stella immediately faced backlash from group members. Stella explains that the ``moderators DM’d me and said that I was like a wuss and that I can't take it. And that if I'm so mentally unwell I shouldn't be on the internet.’’ 
In response, Stella created a splinter group (i.e., a new version of the community) with the same purpose and name except for the addition of ``2.0'' in the title (to denote a new and improved version). Stella explains,
\begin{quote}
``I made a separate group and gave it a similar name to the first one. Everyone found that within five minutes of it being LIVE. I already had requests for people that wanted to do moderation with me. Because they have backgrounds that are similar. So, if there was triggering content, they would be able to read it instead of me.’’
\end{quote}
While proactive, Stella’s approach is labor-intensive which can be challenging for Facebook Group moderators due to the volunteer nature of these roles. Stella’s experience underlines the need for more scalable solutions in online communities, especially for those who lack the time or resources to enforce trigger warnings. Cameron, as well as other participants in the study, values the dedication of administrators like Stella. Cameron wishes for a feature that would provide immediate insight into the moderation of an online group before they decide to join. Cameron explains, 
\begin{quote}
``If I were to come across a group that had a rating [of] a one out of five, that's…based on moderator responses and resolutions…I could have a little bit more caution [and] be like, `You know what, I'm not going to join this group, because either the moderators aren't \textit{able} to resolve issues or they're not \textit{trying} to resolve issues.’’
\end{quote}
In this quote, Cameron describes a tool that would not only save time but also help them avoid personal-communal gaps, which can be detrimental to their online identity reconstruction. In the discussion section, we expand on how this type of feature would help make online communities be more trauma-informed by making them more transparent and empowering for users.

\subsection{Gaps Leading to Online Identity Reconstruction}
\label{sec:gapsoffline}
In this section, we show different ways in which people can glaze over their identity fractures. 
\subsubsection{The emergence of identity gaps} We find evidence of fractures leading to online identity gaps. Eloise, an IPV survivor, explains, ``the people that know me in my everyday life look at me and say, `That's not consistent with the version of you that I know’ or `that's very different from the person that you are now, things have changed,’ [leading them to ask] `what changed?’ I don't have to go through that whole backstory when it's an online community.'' These changes in Eloise’s enacted identity—the way she behaves and interacts—do not align with the expectations of those who knew her before the trauma, reflecting an enacted-personal gap. Eloise's mention of not wanting to delve into her ``whole backstory’’ suggests that recounting the trauma and its subsequent impact on her identity is emotionally taxing. This indicates a personal-relational gap. To avoid the discomfort of explaining herself, Eloise turns to online communities as she doesn't face the same level of scrutiny (or need to explain the changes in her behavior) in these spaces. Not only can one's immediate circle be unable to understand one's trauma but they can also be unwilling to provide support. Daisy explains, ``In reality, your friends and family don't want to talk about it.’’ Lysa's experience as a survivor of assault echoes Daisy’s observation. In the quote below, Lysa describes the  personal-communal gap she encounters when confiding in a fellow military spouse about her assault, which happened while living overseas where her husband was stationed. Lysa shares,
\begin{quote}
``I was sexually assaulted by a local, and I reached out to another military spouse. Like, `Hey, I'm having a lot of trouble. I don't want to leave the house, I'm having all these issues.’ And her first response was, `Well, he didn't actually penetrate you. So, is it really that big of a deal’…Anytime I talk about it, it was, let's downplay it. Let's not make too much of a big deal of it, because your hysterics are going to ruin your husband's career’’
\end{quote} 
In this excerpt, Lysa’s attempt to seek support is met with rejection. The response she receives—``Well, he didn't actually penetrate you. So, is it really that big of a deal?’’—contrasts her own perception of the incident and its effects. This reaction not only minimizes her trauma but also presents a gap between her personal identity and the communal norms of her military spouse community. In this subculture, Lysa describes how silence can often prevail to avoid professional repercussions for their partners. This lack of support ultimately led Lysa to reconstruct her identity in a true crime podcast community. Lysa explains how the members of this online group normalize challenging topics that ``people don't want to bring up in everyday conversations.’’ The readiness of members to provide peer support helps Lysa integrate fractures that are``not necessarily nice’’ or easy for her inner circle to hear about. 

Pat, a survivor of assault, experienced a personal-relational gap when her family and friends ascribed her with the label of victim. She explains, ``They looked at me with pity instead of looking at me [as someone] crying for help. I am not somebody who's ever wanted to be pitied for those things that I've gone through. I've only ever wanted to relate to people…So, if you can find the people that listen, and sometimes doing it online, people will listen. I think that's a really valuable thing.’’ By engaging with others online, who can relate to her experiences, Pat is able to receive help without feeling shame. 

\subsubsection{Alternative safe space} \label{sec:altspaces} Olive, a survivor of harassment, was severely bullied in childhood due to their weight. Today, this part of Olive's enacted identity is celebrated in online groups reflecting identity intersections. In their case, plus-size fans of MyFavoriteMurder, a true crime podcast. Olive and other participants in the sample describe these groups as ``spin-offs,'' ``splinters,'' or ``crossovers.'' Olive explains, ``One [group] is [called] `Stay Fat and Don't Get Murdered' and the other is `Plus-size Murderinos’…We're a community that boosts each other up. But we will make little twisted comments about stuff. And it's really nice, we are a great little group.’’ The way Olive describes their groups, as a collective that ``boosts each other up,’’ speaks to the peer support inherent in these spaces. This support allows Olive to reframe aspects of their enacted identity, previously a target of shame, into attributes that are embraced by others. Eleanor also finds acceptance online as she navigates her recent divorce and confronts pressures to have children, reflecting a personal-communal gap. Eleanor shares, ``I found a really great crossover [group], MyFavoriteMurder and child-free, where I finally feel accepted and normal for not having children because there's a lot of pressure to be married and have kids and that was really just a great place to be online finally.’’ In this community, Eleanor feels ``accepted and normal’’ for her life choices. This acceptance is crucial for Eleanor’s identity reconstruction as it allows her to incorporate her fractures in a more integrated identity.

Courtney, introduced in section \ref{sec:fractures}, is part of a Facebook Group for osteology hobbyists.\footnote{The scientific study of human bones.} In this group, discussions about trauma naturally surface among members despite the group's primary focus being about the scientific study of bones. She explains, ``When you join those sorts of groups, things come out. And it's almost like cognitive grouping but for trauma survivors. We all just sort of cling together and we're like, ‘Oh, wow, this is really neat. But do you think it's because of [trauma]?’’’Courtney notes that survivors within the group ``cling together,’’ suggesting that this community offers a sense of belonging. In her quote, Courtney shares how group members discuss if their interests in collecting bones is due to their traumatic experiences, exemplifying how peer support emerges in alternative safe spaces like a group for osteology hobbyists. 

Katrina, a rape survivor, prefers to reconstruct her identity through private conversations on Facebook Messenger, which differs from Courtney's approach of finding community in a public Facebook Group. Katrina's preference for private interactions is relevant in the context of rape which is often associated with stranger-perpetrated violence or non-consensual acts involving overt physical force. Sexual assault committed within intimate relationships, however, is not always universally recognized. This lack of recognition is often rooted in prevailing societal norms and misconceptions. Consequently, participants who experience this trauma describe grappling with a personal-communal gap as they are often confronted with whether their experience aligns with commonly understood definitions of assault. This personal-communal gap can lead survivors like Katrina to use features like private messaging on Facebook. Katrina explains, 
\begin{quote}
``I wouldn't say I've ever posted my experience on a main platform. But I have been raped. And I've definitely reached out to other people that have shared their story and…[have] privately messaged them, `Hey, this happened to me. Do not blame yourself. I spent nine months not even believing that I was raped’...So anytime I see women struggling with guilt, I feel so compelled to reach out because I blamed myself. I didn't even tell my friends about it because I was so embarrassed.’’
\end{quote}
Katrina's choice to use private messaging as a means of connecting with others underscores the need for secure, private spaces in online environments. These spaces allow survivors like Katrina to express themselves and seek peer support, a trauma-informed principle, without the fear of public judgment or scrutiny. Katrina’s experience stresses the importance of designing support systems within online communities that are sensitive to these identity gaps which we expand on in section \ref{sec:design_rec} on design recommendations.

\section{DISCUSSION} 
In this section, we reflected on how the findings relate to earlier work on identity reconstruction in online communities. To answer RQ1 specifically, we reflected on the importance of extending the CTI to incorporate the affordances of online spaces and Goffman's dramaturgy framework to include improvisation. In association with these theoretical themes, we will introduce design recommendations through the lens of trauma-informed design to answer RQ2.

\subsection{Theoretical Implications}
In this section, we situate our findings in earlier theoretical frameworks on identity \cite{goffman_presentation_1956} and extend CTI \cite{hecht_communication_2005} to include platform design. In section \ref{sec:balance_safety}, we contextualize our findings through the lens of trauma-informed design and conclude with a discussion of how algorithmic systems can trap users in grief bubbles, impeding their identity reconstruction.

\subsubsection{Introducing the Sociotechnical Model of Identity Integration}
\label{sec:model}
\begin{figure*}[ht]
  \centering
  \includegraphics[scale=0.5]{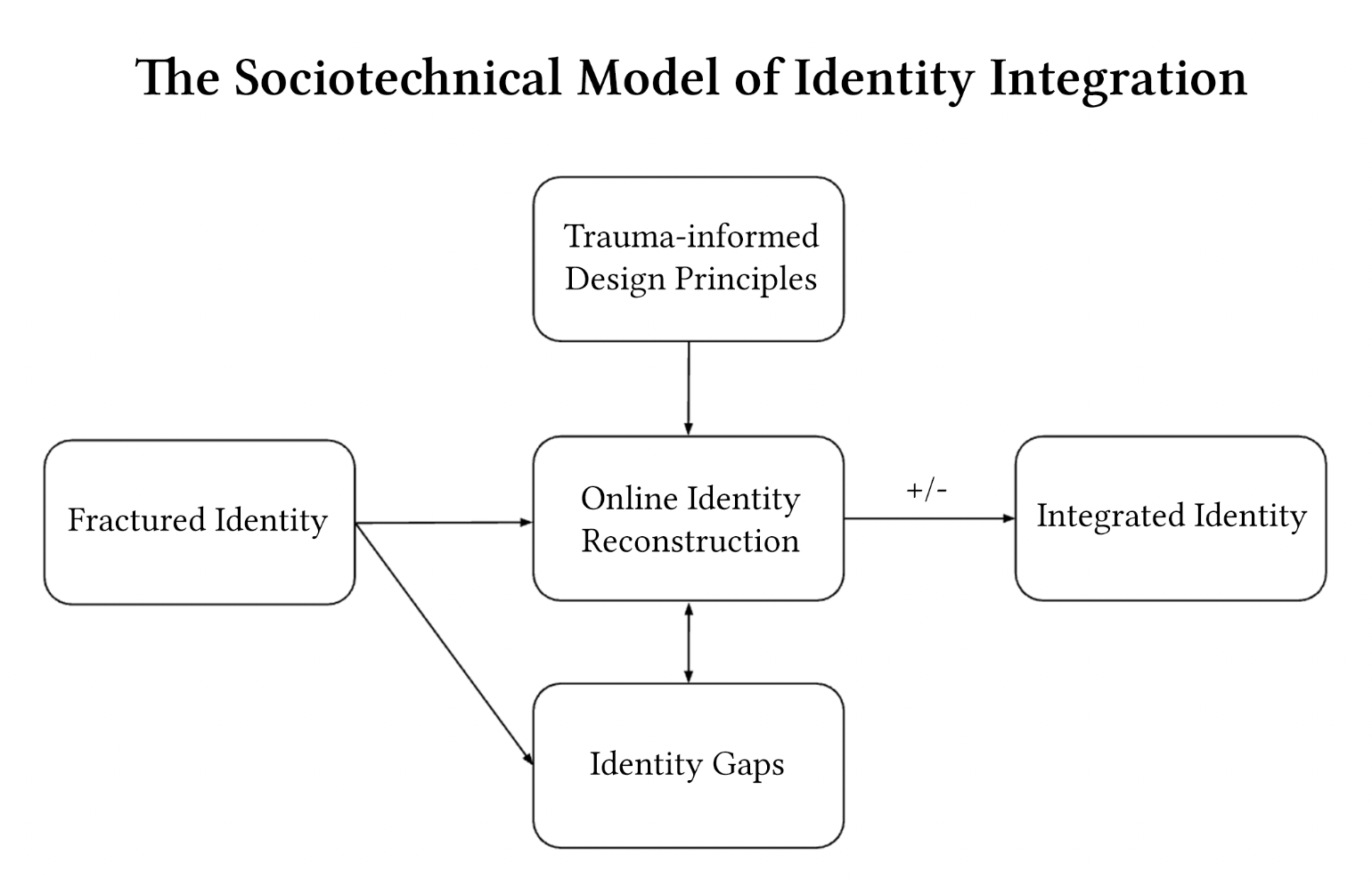}
  \caption{The Sociotechnical Model of Identity Integration}
  \Description{This model illustrates how the process of online identity reconstruction acts as a mechanism to incorporate fractured identities into one's self concept.}
  \label{fig:model}
\end{figure*}

Our findings provide empirical evidence for a sociotechnical model (Figure \ref{fig:model}) that 
illustrates the process of reconstructing identities online after trauma. The model begins with a fractured identity (reflecting how trauma affects an individual’s perception of self) and extends to a bidirectional relationship between identity gaps (i.e., contradictions among identity frames) and online identity reconstruction (i.e., redefining one’s self-concept in communicative processes). Our data reveals that fractures can either prompt identity gaps (e.g., personal-communal, personal-relational) and online identity reconstruction can itself generate new gaps (e.g., personal-communal, enacted-communal). 

We find that trauma-informed design principles significantly influence whether these processes 
lead to identity integration (i.e., weaving fractures into a renewed sense of self). 
Notably, one positive pathway emerges as a \textit{Kintsugi-based approach} to integration, where survivors \textit{glaze} their fractures rather than concealing them, resulting in an identity that acknowledges the trauma while fostering growth. Conversely, misaligned design decisions can create barriers and deepen fractures (e.g., grief bubbles). In this sense, the sociotechnical pathway can yield a positive (``+'') outcome, where survivors move toward an integrated identity, or a negative (``-'') outcome, where fractures widen or new ones emerge.

\subsubsection{The function of online communities in identity reconstruction} In addressing RQ1, we find that trauma-informed online communities are pivotal to the process of online identity reconstruction. Building upon prior work \cite{Andalibigarcia2021, ammari_crafting_2017, andalibi_announcing_2018, haimson_trans_2020, Haimsonoli2018, Semaanetal2016, Semaanetal2017, ammari_accessing_2014, ammari_thanks_2016}, we confirm that these digital spaces can provide an avenue for survivors to address their fragmented sense of self. Eloise's experience in \ref{sec:gapsoffline} exemplifies how online interactions can alleviate gaps between one's personal and enacted frames. The examples in \ref{sec:gapsoffline} demonstrate how online communities can enable survivors to find support and understanding often lacking in their immediate social circles. Lysa, for example, shares how the same topics that are avoided in face-to-face interactions are often normalized within her online groups, particularly ones that act as alternative safe spaces \cite{randazzo2023}. We also find that online communities can integrate \textit{identity fractures} which we conceptualize as changes in self-perception due to traumatic experiences. The concept of identity fractures elaborates on Janoff-Bulman’s \cite{janoff-bulman_shattered_2002} `the shattered self' by situating fractures on Hecht's \cite{hecht1993} communication frames of identity. In Figure \ref{fig:IdentityIntegration}, identity fractures are painted with gold to symbolize the potential of these communities to glaze fractures, echoing the Japanese art of Kintsugi \cite{koren1994wabi}. Similar to Kintsugi, these spaces can provide a supportive environment where fractures are embraced and integrated into one's self-concept.

Focusing on the enduring nature of identity fractures \cite{janoff-bulman_shattered_2002, herman1992}, we used sociomateriality as a lens to incorporate the nature of online communities and design of social media platforms into CTI \cite{hecht1993, jung2004elaborating} which acknowledges the multiplicity of identities (e.g., personal, communal, relational). This allows us to better understand identity reconstruction associated with enduring identity fractures resulting from traumatic experiences. 

Our findings also reveal how online communities can reflect the enacted identity of trauma survivors through algorithmic interactions and group norms. Future work should explore this evidence alongside prior work on sociomateriality \cite{barad2007meeting, van2011feminist, butler2013contingent}. Extending Hecht's \cite{hecht1993, kuiper2023bridging} CTI, we argue that the design of online spaces can be a critical aspect of one's enacted identity. In the context of HCI, personal-enacted gaps are discrepancies between an individual's self-concept and how their identity manifests and is interpreted in online environments. We expand on how personal-enacted gaps can emerge as a result of algorithmic interactions in section \ref{sec:bubble}.

\subsubsection{Moving beyond Goffman} This study challenges Goffman's \cite{goffman_presentation_1956} assumption of identity salience for users reconstructing their identities after trauma. Our findings contradict Goffman's \cite{goffman_presentation_1956} dramaturgy framework which suggests a degree of control in how individuals present themselves using social scripts. This perspective, however, neglects the identity reconstruction experiences of trauma survivors. Survivors like Lars in section \ref{sec:fractures} indicate an unscripted and evolving process of identity formation that reflects \textit{improvisation} more than a meticulously planned performance \cite{goffman_presentation_1956}. To extend Goffman's work, we argue that improvisation\footnote{Improvisation can be considered a form of identity play \cite{dym2019coming}} is critical to the identity reconstruction process for trauma survivors.

We also provide evidence for the practical use of Hecht’s \cite{hecht1993} CTI for the examination of online identity reconstruction.\footnote{To the best of our knowledge, few studies use CTI as a framework to consider the role or design of communication technologies with the exception of \cite{riemen2019, brooks2016communication}.} CTI not only incorporates Goffman's \cite{goffman_presentation_1956} work on identity performance (within the enacted identity frame) but also allows scholars to examine how users navigate multiple identity frames simultaneously. As our model demonstrates, the application of CTI can enable scholars to investigate the gaps and intersections of identity frames \cite{jung2004elaborating, crenshaw2013demarginalizing}. This approach moves beyond Goffman's \cite{goffman2009stigma} limited focus on `spoiled identities' towards a more dynamic understanding of identity and its intersections.\footnote{Goffman's \cite{goffman2009stigma} work on spoiled identity acknowledges how different social positions (e.g., race, class, gender) can compound or magnify stigma without considering their intersections identities \cite{crenshaw2013demarginalizing}.} Goffman's \cite{goffman_presentation_1956} work assumes identity salience which misses the multifaceted nature of identities as CTI's identity frames exhibit \cite{jung2004elaborating, orbe2004negotiating, brooks2016communication, wadsworth2008role, hecht_communication_2005, hecht1993}.  To illustrate this point, consider the example of a waiter at a restaurant. This individual can put on a performance to diners (frontstage) while revealing their true selves (backstage) to colleagues in the restaurant's kitchen. In this scenario, the individual's professional identity (waiter) is most salient as it is being highlighted while other aspects are suppressed. An intersectional approach recognizes that many aspects of one's identity, like in Lysa's case, being an assault survivor and a military spouse, are inextricably linked and cannot be suppressed or turned off at will. The waiter's identity is not merely confined to their professional role. They could be a person of color navigating a predominantly white work environment, a female employee in a male-dominated industry, or someone with a hidden disability that influences their interaction style. All these aspects interlock to create a more complicated social stage, where the waiter's behavior isn't just shaped by their role as a service provider, but also by the multifaceted nature of their personal identity. Thus, Goffman's \cite{goffman_presentation_1956} perspective is limited because it does not consider other aspects of a survivor's identity, which can sit at intersections, be innate to who they are, and shape their online communication with others. Therefore, to design trauma-informed platforms for identity reconstruction, we argue that scholars should adopt multifaceted identity theories such as Hecht's \cite{hecht1993} CTI which better reflects the multiplicity of identity reconstruction in online environments. As a result, we call upon HCI scholars to move beyond Goffman and adopt more multifaceted frameworks of identity, enabling a more in-depth understanding of online identity reconstruction, especially for fractured identities.

\subsubsection{Balancing safety and support}\label{sec:balance_safety} Our findings provide more insight into the debate on trigger warnings. Luisa’s moderation approach (no-trauma-dumping) in section \ref{sec:triggerluisa} resonates with the trauma-informed principle of user safety \cite{chen_trauma-informed_2022, Scott2023}. Banning trauma disclosures can limit the group's capacity to function as an alternative safe space \cite{randazzo2023}, which counteracts the trauma-informed principle of peer support. Recall Courtney's experience in the osteology group in section \ref{sec:altspaces}. This group's ability to serve as an alternative safe space, despite their main purpose being about bone collection, aligns with the recommendations of Ammari et al. \cite{ammari_moderation_2022} on the importance of flexible community governance. However, maintaining flexibility between openness and safety, as Andalibi et al. \cite{andalibi_announcing_2018} argue, can be challenging in online communities. Stella’s method of implementing trigger warnings strives for such a balance but it is a laborious task \cite{boysen2017evidence} that puts moderators at risk of retraumatization \cite{randazzo2023}. Online communities need a more practical and sustainable solution which allows survivors to access trauma narratives (a crucial communicative mechanism for identity reconstruction; \cite{herman1992, randazzo2023, dunn2010judging}) without triggering harmful memories.  

\subsubsection{Algorithmic identities} \label{sec:algorithmic_identity} Prior work on filter bubbles primarily addresses the spread of misinformation and polarization (e.g., \cite{pariser_filter_2011,sunstein2018republic}). In this work, we introduce the concept of \textit{grief bubbles} which describes algorithmic systems guiding users towards and organizing them into groups predominantly focused on the traumatic aspects of their identities. This phenomenon represents a specific manifestation of filter bubbles, where algorithms inadvertently amplify and reinforce a singular focus on grief and trauma within online spaces. Recall Stavros's experience in section \ref{sec:gapsonline}, in which grief bubble act as both a comfort zone and a trap. While grief bubbles can initially serve as a space for advocacy or venting, the reinforcing nature of these spaces can eventually lead to emotional exhaustion and identity reductionism (i.e., identities being reduced to singular markers; \cite{garrett2002personal}). As Stavros explained, the emotionality of his content recommendations helped fuse his enacted identity as an advocate to his personal identity as a survivor. This finding extends Randazzo and Ammari’s \cite{randazzo2023} work by demonstrating how algorithmic mirrors can encapsulate users within their trauma or what Gillespie \cite{gillespie2014relevance} refers to as recursive loops. In section \ref{sec:gapsonline}, Pat’s experience demonstrates how grief bubbles can lead to symbolic algorithmic annihilation \cite{Andalibigarcia2021}, impeding a survivor’s online identity reconstruction by reducing the complexity of their identities. In section \ref{sec:design_rec}, we provide algorithmic design recommendations that pull users out of grief bubbles and push them toward identity integration.

Algorithmic sorting is posing challenges for users who need access to peer support. Leveraging Hecht's \cite{hecht1993} CTI helps us understand that just because someone is a peer in one experience (both rape survivors) does not mean that they are a peer in another (part of the same generation or culture). Recall that Daisy received indirect feedback \cite{randazzo2023} that started with `Back in my day.’ This phrase hints at a generational approach to coping that might not reflect modern practices, and thereby limits users' access to support. User experience is increasingly being shaped by algorithms \cite{gillespie2014relevance}, the absence of trauma-aware algorithmic sorting can hinder the identity reconstruction process. On platforms like Reddit and Facebook, where certain comments can be algorithmically promoted, the lack of sophistication in distinguishing between different tones can exacerbate feelings of isolation or misunderstanding amongst survivors. This is especially important in discussions around sensitive subjects, where the path to an integrated identity can be disrupted due to inadequate algorithmic design.

\subsection{Design Recommendations}
\label{sec:design_rec}
In this section, we present design recommendations based on the findings in section \ref{sec:findings}. Chen et al. \cite{chen_trauma-informed_2022} argue the broader benefits of trauma-informed design, extending beyond those directly impacted by trauma. We focus on design recommendations to release users from grief bubbles, discuss a context-aware messenger, and highlight moderation transparency. 

\subsubsection{Releasing users from grief bubbles} To release survivors from grief bubbles, as we discuss in section \ref{sec:bubble}, we recommend the design of more sensitive algorithms. This requires incorporating trauma-informed design principles, such as user safety, to provide a more comprehensive understanding of how algorithmic systems impact users' informational environments and their emotional landscapes. Therefore, we recommend a shift towards a more sensitive algorithmic design that is trauma-aware. Existing algorithms often reinforce emotional states by promoting similar content, trapping users like Stavros in a loop of angered advocacy. A trauma-aware algorithm can recognize patterns related to trauma and subtly diversify the content displayed to the user, mitigating the risk of re-traumatization. We also recommend the use of prompts that encourage users to interact with a broader spectrum of content. Inspired by Pat in section \ref{sec:gapsonline}, we argue that prompts can gently guide users out of grief bubbles, hindering the path toward identity integration. In effect, the grief bubble can be pierced by offering members different content from trusted sources \cite{nguyen2018escape} in order to eventually allow them to engage in wider discursive spaces \cite{maciag2018discursive}. 

\subsubsection{Context-aware messenger} Katrina's experience sheds light on the importance of affordances in the digital space and the contextual relevance of identity fractures. Katrina’s decision to privately message group offers her a sense of security, echoing the trauma-informed principle of user safety.  However, in digital environments like Facebook Messenger, these conversations can potentially expose users to emotional risks. Therefore, we recommend a design that separates conversations within group contexts from other general messaging interactions. This would ensure that discussions related to traumatic experiences remain within the safe confines of the relevant group, thereby safeguarding users' emotional well-being. Our findings highlight how a lack of context-aware design in digital environments can hinder the process of identity integration. For survivors in the sample, separating sensitive conversations in digital spaces helps to uphold the trauma-informed principle of user safety and protect their emotional well-being as they reconstruct their identities online.

\subsubsection{Moderation transparency} \label{sec:mode_transparency} The role of moderators in shaping online communities cannot be understated \cite{takahashi_potential_2009} as we illustrate in section \ref{sec:mod}. Still, social media platforms can be more transparent about moderation practices through the inclusion of a metric which we refer to as \textit{Moderation Response}. This visible metric can be used as a tool by potential members to assess the \textit{presence} and \textit{effectiveness} of a community's governance before joining a group. Attaching a public metric to online communities can also help platforms hold group administrators accountable for the moderation of their group. We recommend displaying this metric similarly to `response rates’ often found in the customer service section of Facebook Business pages.

Our research supports Randazzo and Ammari's \cite{randazzo2023} call for social media platforms to disclose precise metrics on the efficacy of their trauma-care tools. As we discussed in section \ref{sec:mod}, Reddit provides limited insight into the actual usage of RedditCares which does not adhere to the trauma-informed principle of transparency and trustworthiness \cite{chen_trauma-informed_2022, Scott2023}. Also, enlisting users to flag posts (that express intentions of suicide or self-harm) raises concerns about Reddit’s practice of using community members for unpaid moderation labor \cite{gillespie2018custodians}. This arrangement positions users as frontline moderators without the transparency or accountability typically associated with such a role. To improve the design of this tool, we propose that platforms provide at least group administrators with detailed usage metrics of RedditCares within their communities. In doing so, administrators can gain a clearer understanding of whether and how trauma-care tools are being abused. This information can empower group administrators to adjust community norms and guidelines effectively, helping to foster a safer environment for survivors in the process of reconstructing their identities online. In doing so, these platforms can uphold the principles of trauma-informed design by making moderation practices more transparent and trustworthy.

\section{LIMITATIONS AND FUTURE WORK}
While our study offers valuable insights, we recognize its limitations as opportunities for future exploration and growth. We recruited participants from online forums that are trauma-adjacent or focus on trauma recovery. Exploring a broader range of communities can potentially enrich the understanding of supportive online environments. We also recommend that future studies apply experimental methods to test the influence of the moderation visibility metric on community participation and other user behaviors. Future work can also investigate the concept of grief bubbles which can deepen our understanding of emotional well-being and identity reconstruction in algorithmic systems. Additionally, we advocate for the use of participatory design methods when implementing our design recommendations, recognizing the value of involving users in the development process. Thus, this approach aligns the final design with the lived experiences of the participants \cite{kensing1998participatory}.

Another limitation of this study is that our sample skews heavily white and female. The challenge of recruiting  minorities and marginalized populations in mental health research has been studied earlier \cite{brown2014barriers}. Future work should oversample members of minority groups to better understand their design needs. 

\section{CONCLUSION}
This study investigates the complexity of reconstructing identities online after trauma. Our findings reveal a process model which demonstrates the sociotechnical pathways from a fractured to integrated identity. Our research highlights how online communities, while offering support, can also unintentionally intensify the challenges faced by trauma survivors due to a lack of trauma-informed design principles. The study also sheds light on the role of algorithms in shaping these experiences, often creating grief bubbles that can isolate or emotionally overwhelm users. In response to the findings, we provide trauma-informed design recommendations that address the multifaceted nature of identities, aiming to support users in reconceptualizing their sense of self after trauma.

\section{Acknowledgments}
Thank you to the participants for confiding in us and to the administrators and moderators who allowed us to recruit from their online communities. Special thanks to the reviewers who took the time to provide us with detailed feedback. 

\bibliographystyle{ACM-Reference-Format}
\bibliography{software}

\end{document}